\documentclass[twocolumn,aps,prl,floatfix,superscriptaddress]{revtex4-1}
\usepackage{amsmath,amssymb}
\usepackage{graphicx,float}
\usepackage{times,txfonts}
\usepackage{color}
\usepackage{multirow}
\usepackage{bbold}
\usepackage{color}
\usepackage{soul}
\usepackage{hyperref}
\usepackage{times,txfonts}
\usepackage{natbib}
\usepackage{nicefrac}
\usepackage{blindtext}
\usepackage[export]{adjustbox}
\usepackage{mathrsfs}
\usepackage{textcomp}
\usepackage{blkarray}
\usepackage{multirow}	

\newcommand{\op}[1]{\widehat{#1}}
\newcommand{\bra}[1]{\left\langle #1 \right|}
\newcommand{\ket}[1]{\left| #1 \right\rangle}
\newcommand{\braket}[2]{\left\langle {#1{\left| \vphantom{#1 #2} \right.} #2} \right\rangle}

\renewcommand{\epsilon}{\varepsilon}

\def\VR{\kern-\arraycolsep\strut\vrule &\kern-\arraycolsep}
\def\vr{\kern-\arraycolsep & \kern-\arraycolsep}
\usepackage{sistyle}
\usepackage{soul}
\definecolor{lightblue}{RGB}{185,210,248}
\sethlcolor{lightblue}

\begin{document}

\title{High-Dimensional Quantum Certified Deletion}

\author{Felix Hufnagel}
\email{fhufn079@uottawa.ca}
\affiliation{Nexus for Quantum Technologies, University of Ottawa, Ottawa, K1N 6N5, ON, Canada}

\author{Anne Broadbent}
\affiliation{Nexus for Quantum Technologies, University of Ottawa, Ottawa, K1N 6N5, ON, Canada}
\affiliation{Department of Mathematics and Statistics, University of Ottawa, Ottawa, Ontario, K1N 6N5 Canada}

\author{Ebrahim Karimi}
\affiliation{Nexus for Quantum Technologies, University of Ottawa, Ottawa, K1N 6N5, ON, Canada}

\begin{abstract}
Certified deletion is a protocol which allows two parties to share information, from Alice to Bob, in such a way that if Bob chooses to delete the information, he can prove to Alice that the deletion has taken place by providing a verification key. It is not possible for Bob to both provide this verification, and gain information about the message that was sent. This type of protocol is unique to quantum information and cannot be done with classical approaches. Here, we expand on previous work to outline a high-dimensional version of certified deletion that can be used to incorporate multiple parties. We also experimentally verify the feasibility of these protocols for the first time, demonstrating the original 2-dimensional proposal, as well as the high-dimensional scenario up to dimension 8.
\end{abstract}	

\maketitle

\noindent\textit{Introduction --} In the current climate of remote services and mass data storage, the ability to know if someone has deleted information that you have sent to them or asked them to hold onto for some period of time may be as important as communication security. Going forward, a verifiable proof that a company has deleted personal data may be integral to our continued faith in cloud storage and data-collecting companies. The inability to make copies of a general quantum state, described by the no-cloning theorem~\cite{park1970concept,wootters1982single}, is a fundamental aspect of many proposed quantum technologies, including quantum key distribution (QKD)~\cite{bennett1984quantum} and blind quantum computing~\cite{broadbent2009universal}. Such technological solutions use the physical properties of quantum mechanical systems to gain a security advantage over the previously used digital approaches. QKD has become a frontrunning solution to secure communication in a future where access to quantum computing resources will render current security protocols useless. This field has received significant attention both from the theoretical physics and mathematics community which has developed new protocols and security proofs to optimise the original communication protocol proposed in 1984~\cite{bennett1984quantum}, and from experimental physics which has pushed the bounds of what can actually be achieved with fibre channels~\cite{sit:18,rosenberg2007long,gobby2004quantum}, underwater channels~\cite{bouchard2018underwater}, and free-space channels~\cite{Sit:17,schmitt2007experimental} linking line of sight stations within cities and from ground to satellite~\cite{Yin:2017}. Furthermore, this technology is beginning to enter the commercially available phase of development with a few pioneering companies such as ID Quantique, Toshiba, and MagiQ Technologies Inc., to name a few, producing specific-use products. Eventually it is expected that a large infrastructure of quantum communication channels will be used to allow secure communication around the world. On the back of this infrastructure one can begin to consider other applications for the quantum channels such as blind quantum computation, quantum money, and the quantum internet etc.

A more recent proposal, again resting on the no-cloning theorem for quantum states, has defined a protocol for certified deletion~\cite{broadbent2020quantum}. The certified deletion protocol allows for a receiving party to guarantee that information sent to them has in fact been deleted and that no copy has been held by the receiving party. Such a protocol is not possible in the world of digital communication where a person can always hold on to the raw bits that have been sent, and recreate a copy. This has motivated protocols using certified deletion for data security and privacy, software licensing, and new public encryption schemes using quantum resources~\cite{garg2020formalizing,poremba2022quantum,bartusek2022cryptography}. There have also been similar ideas around proving erasure of quantum data stored at some remote location~\cite{coiteux2019proving}. This allows for one party to store a backup of their data at some location while being able to request a proof that this data is deleted at a later date. Here, we experimentally demonstrate the certified deletion protocol proposed in~\cite{broadbent2020quantum}, using the orbital angular momentum degree of freedom of photons. In addition, we extend the protocol beyond the qubit, aiming to develop the certified deletion protocol to utilise high-dimensional quantum communications systems.
\begin{figure}[H]
	\begin{center}
	\includegraphics[width=1.0\columnwidth]{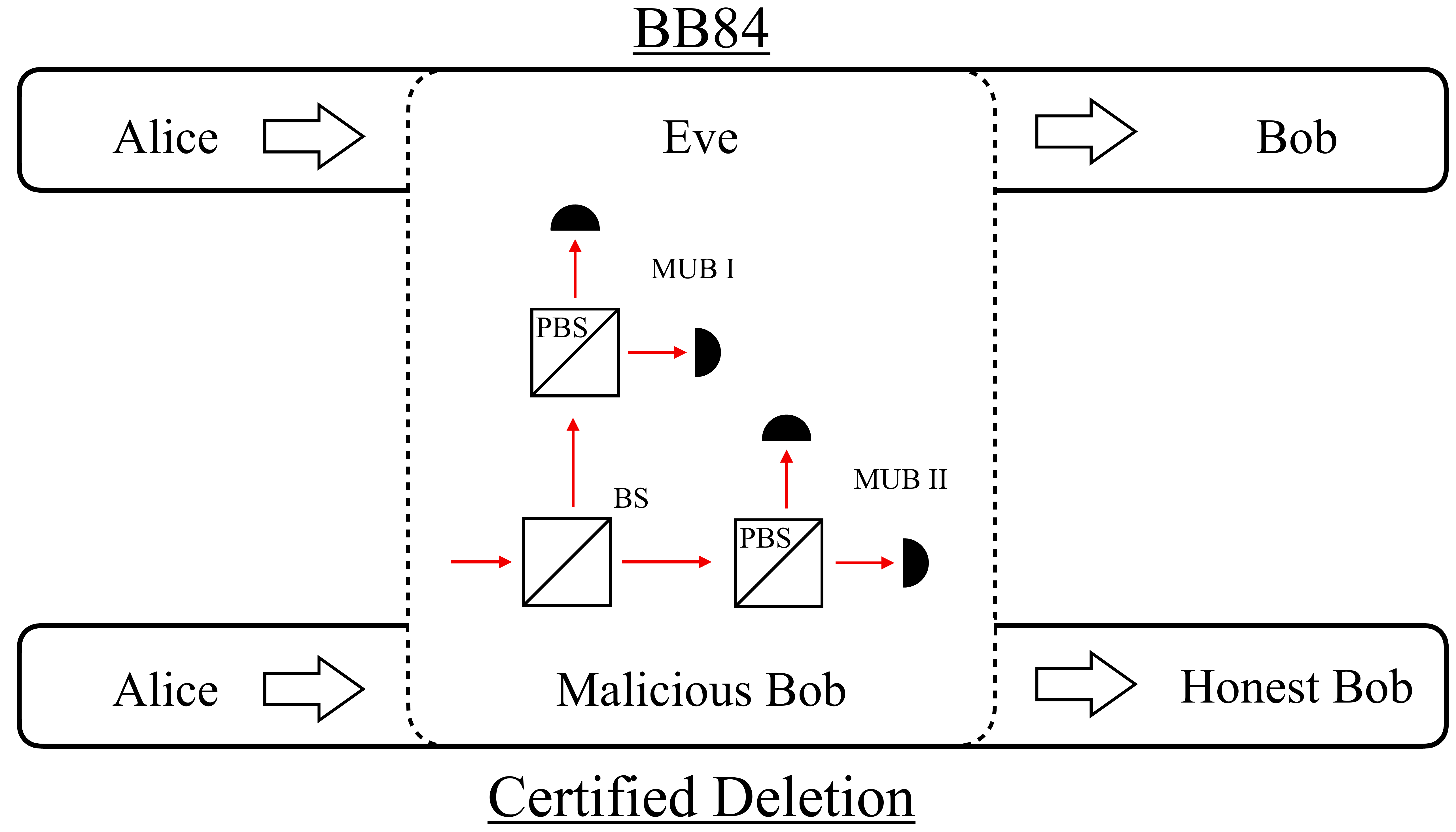}
	\caption{Here we describe the relationship between BB84 and certified deletion. In the certified deletion protocol the malicious Bob, who is trying to determine both the deletion key and the secret message, maps onto Eve in the BB84 protocol. For the security proof, we can describe the certified deletion protocol as Alice sending the quantum state to Malicious Bob and then on to the Honest Bob.}
	\label{fig:BB84}
	\end{center}
\end{figure}
\noindent\textit{Protocol --} A fundamental aspect of certified deletion is shared with QKD, that of conjugate coding using mutually unbiased bases (MUBs). By encoding information into conjugate bases, we are able to take advantage of the quantum no-cloning principle, leading to mathematical limits on the eavesdropper's abilities in QKD or to Bob's ability to \emph{both} convince Alice of deletion \emph{and} extract information about the message in certified deletion. These limits provide us with a guideline of experimental error thresholds which we must keep our quantum system below to allow for secure communication, and consequently certified deletion. The specifications of a particular protocol will dictate the form of the security proofs and thus will change the error thresholds for different protocols. These specifications include the types of states used, the dimensionality of those states, as well as the choice of measurements that will be performed. Different protocols will have advantages in terms of improved message rates or improved tolerance to errors, but will also ultimately depend on the practical ability to create certain quantum states and perform complex measurements. Let us now extend the certified deletion protocol to the high-dimensional vector space. We will begin by using concepts from high-dimensional QKD schemes that have been developed as extensions to the original BB84~\cite{bechmann2000quantum}. 

A generalised quantum measurement can be given as set of linear operators on a quantum system $A$ denoted by $\left\{\op{M}_A^x\right\}$, satisfying $\sum_{x}\left(\op{M}_A^x\right)^{\dagger}\left(\op{M}_A^x\right) = \mathbb{1}_A$, where, $x \in\left\{1,2,\ldots\right\}$, and~$\dagger$ stands for the conjugate transpose. When $\sum_{x} \op{M}_A^x = \mathbb{1}_A$, these operators are known as a positive-operator valued measure (POVM). MUBs are a particular class of POVMs whose defining feature is that the overlap of two different bases, denoted by $\op{M}^x_A$ and $\op{N}^y_A$, gives $c = max_{x,y} \left\| \sqrt{\op{M}_A^x} \sqrt{\op{N}_A^y}   \right\|^2_{\infty} = 1/d$, where $x,y\in\left\{1,\ldots,d\right\}$, $d$ is the dimension of the vector space, and $|| \op{O} ||_{\infty} = \underset{i}{\max}|o_i|$ is the infinity norm. 

For certified deletion, the string of bits composing the message from Alice are encoded in the computational basis, $\Pi_1 = \left\{ \op{M}^x_A = \ket{\psi_x}\bra{\psi_x}\right\}$, and the deletion key is encoded in the Hadamard basis, $\Pi_2 = \left\{ \op{N}^y_A=\ket{\phi_y}\bra{\phi_y}\right\}$. The choice of ordering for sending in the computational and Hadamard basis is randomised, determined by a random number generator. Thus due to the use of the mutually unbiased bases, Bob can only get information either about the message, by measuring in the computational basis, or the deletion key, by measuring in the Hadamard basis. We show that an attempt from Bob to obtain information about both the key and message is equivalent to an eavesdropper attack on a QKD scheme, and thus, our security proof can use proofs developed for quantum communication. We can draw a picture here relating certified deletion to BB84, Fig. \ref{fig:BB84}. In the certified deletion case we must consider as an adversary a malicious Bob who is trying to determine both the deletion key and secret message, and an honest Bob. The malicious Bob here maps onto Eve in BB84, where Eve is trying to find out information without introducing error, which can be generallized to Eve trying to find out information about one basis without introducing errors into the conjugate basis. To frame it in another way, Bob must provide Alice with the deletion key, meaning he makes each measurement in the Hadamard basis. While Eve tries to determine Alice's states in the computational basis only to gain information on the secret message. 

As in QKD protocols, a certain upper bound is established for the number of errors allowable, here in the proof of deletion Bob provides to Alice. 

The mutual information between Alice and Bob is given by
\begin{equation}
	I_{AB}= \sum_{ij} P(x_i, y_j)\,\text{log}_2\left(\frac{P(x_i, y_j)}{P(x_i)P(y_j)}\right),
\end{equation}
where the $P(x_i, y_j)$ are the probabilities of the outcome $x_i$ for Alice and $y_i$ for Bob, and $P(x_i)$ and $P(y_i)$ are the independent probabilities of each outcome for Alice and Bob. Thus in the high dimensional case, \emph{i.e.}, qudits, we consider a uniform probability of detection errors in Bob's measurement, and can give the mutual information for Alice and Bob in terms of Bob's state fidelity $F$ by,
\begin{equation}
	I_{AB} = \text{log}_2(d)+F\,\text{log}_2\left(F\right)+\left(1-F\right)\,\text{log}_2\left(\frac{1-F}{d-1}\right).
	\label{eq:mutalAB}
\end{equation}
Bob's state fidelity is given by $F = \bra{\psi_i}\op{\rho}_B\ket{\psi_i}$ in the computational basis and by $F = \bra{\phi_i}\op{\rho}_B\ket{\phi_i}$ in the Hadamard basis where $\op{\rho}_B$ is the quantum state received by Bob via the quantum channel. This fidelity also corresponds to the trace of the probability of detection matrix that will be measured in the experimental section to characterise the quantum channel and yields the quantum bit error rate (QBER) through $\text{QBER}= 1-F$.

It has been shown that a limit on Eve's, \emph{i.e.}, malicious Bob's, information can be derived from an uncertainty principle approach which limits the information that can be gained by Eve and Bob from making measurements $\widehat{E}_A$ and $\widehat{B}_A$ respectively on a quantum system \textit{A}~\cite{Cerf2002}. Using the eigenstates $\ket{e_j}$ and $\ket{b_i}$ for $\widehat{E}_A$ and $\widehat{B}_A$ respectively, this limit is given by, 
\begin{equation}
	I_{AB} + I_{AE} \le 2\,\text{log}_2\left(d \max_{i,j} \left|\braket{b_i}{e_j}\right|\right).
\end{equation}
From here, we know that a measurement by Eve in the complementary MUB to Bob will give $\left|\braket{b_i}{e_j}\right|^2=1/d$ and thus $I_{AB}+I_{AE}\le\text{log}_2(d)$. Finally, the proposition $R\ge \text{max}\{I_{AB}-I_{AE},I_{AB}-I_{BE}\}$, detailing the necessary restriction on the mutual information for generating a message, gives the result that we can establish a non-zero message rate for $I_{AB}>I_{AE}$. Here, $I_{ij}$ defines the mutual information between $i$ and $j$, where $A$, $B$, and~$E$ represent Alice, Bob, and Eve, respectively~\cite{csiszar1978broadcast}. We then can determine that we must have $I_{AB}>\text{log}_2(d)/2$ to guarantee a positive message rate. Combining this with Eq.~\eqref{eq:mutalAB}, we reach a lower bound on the fidelity required to give a positive message rate:
\begin{equation}
	F\,\text{log}_2\left(\frac{1}{F}\right) + (1-F)\,\text{log}_2 \left(\frac{d-1}{1-F}\right) < \frac{\text{log}_2 d}{2}.
 \label{eq:fidelity}
\end{equation}
Once it is established that Alice and Bob have a sufficiently high mutual information (fidelity, corresponding to a limit on Eve's information), it has been shown that hash functions can be used to further reduce Eve's information all the way to zero. Though this privacy amplification comes at the expense of reducing Alice and Bob's message length. Equation \ref{eq:fidelity} allows us to determine the maximum tolerable QBER, beyond which point a secret message cannot be established. For example, the error thresholds for dimensions 2, 4, and 8 are 11.00\%, 18.93\%, and 24.70\% respectively\cite{Cerf2002}. 

\textit{High-dimensional photonic states --}  In our protocol, we will use orbital angular momentum (OAM) states to encode our quantum information. OAM of photons is defined by its
\begin{figure}[H]
	\begin{center}
	\includegraphics[width=0.7\columnwidth]{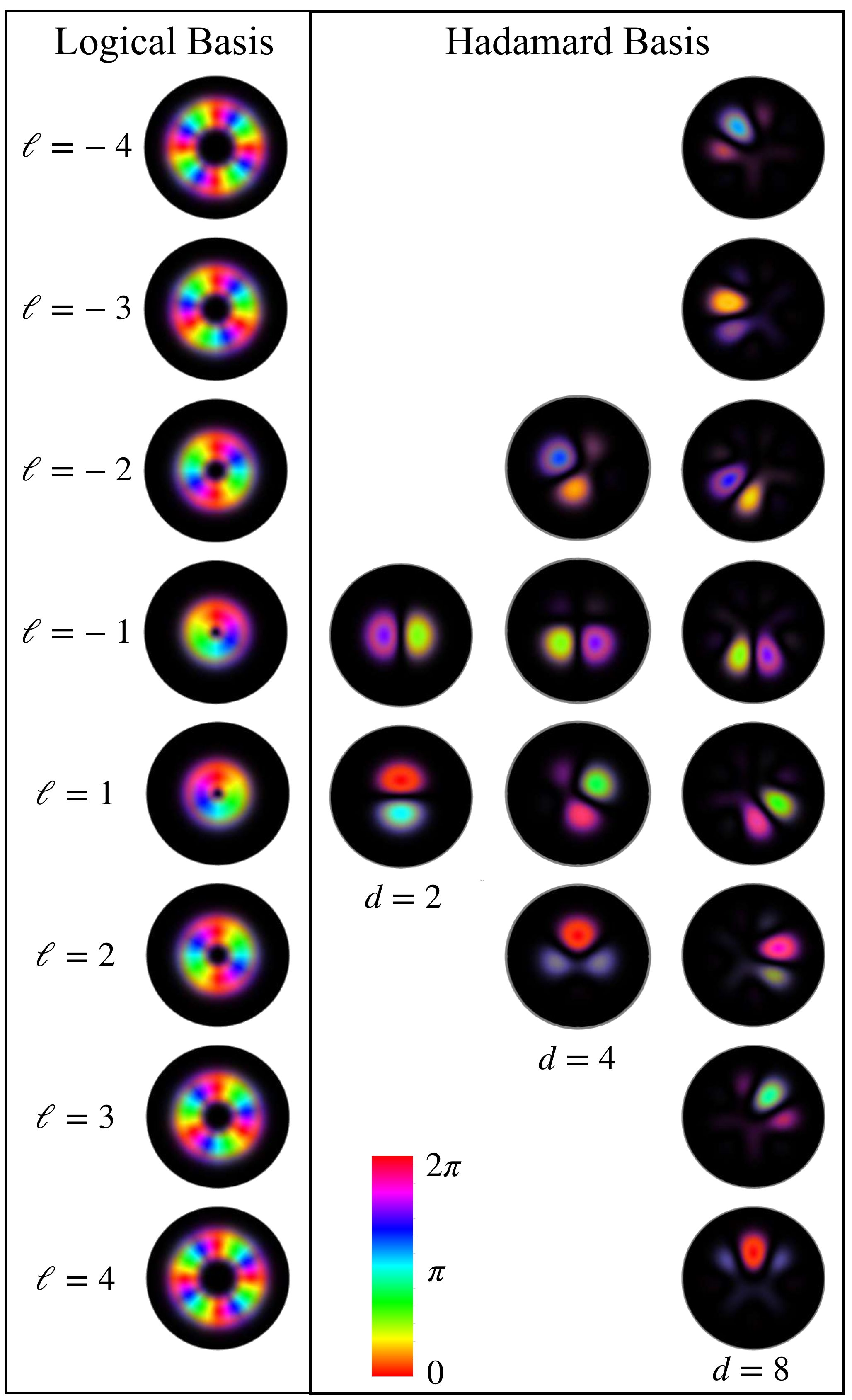}
	\caption{The logical and Hadamard bases for dimensions 2, 4 and 8 are shown. The pure OAM states are shown on the left where $\ell =\left\{-1,+1\right\}$ make up the logical basis for $d=2$; $\ell =\left\{-2,-1,+1,+2\right\}$ for $d=4$; $\ell =\left\{-4,-3,-2,-1,+1,+2,+3,+4\right\}$ for $d=8$. The Hadamard basis states for dimensions 2, 4, and 8, are shown on the right side. The states are plotted with intensity and phase, where the colour ranging from red through the colors and back to red represents a phase of 0 to $2\pi$.}
	\label{fig:modes}
	\end{center}
\end{figure}
\begin{figure*}[!htb]
	\begin{center}
		\includegraphics[width=2.0\columnwidth]{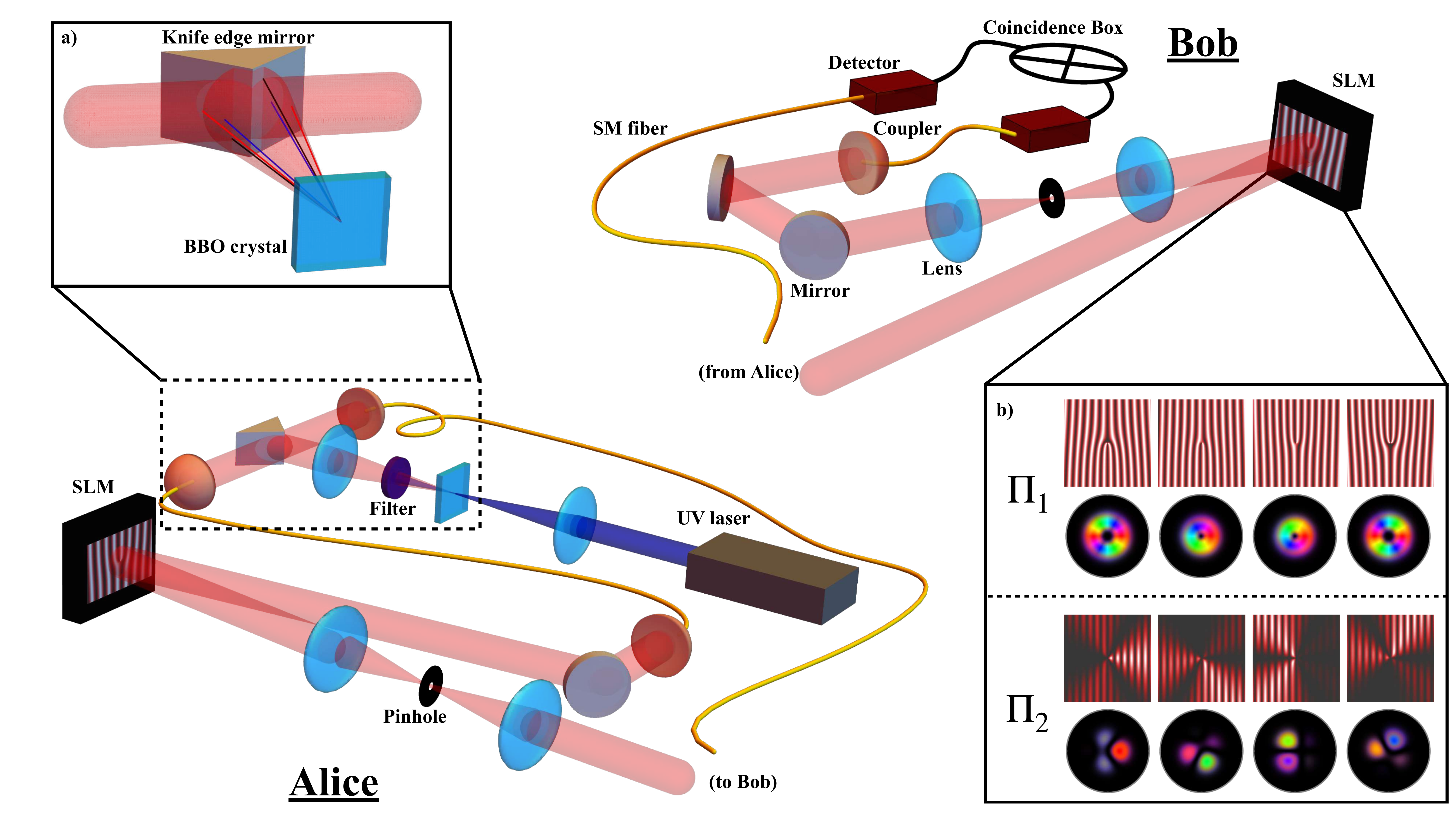}
		\caption{Experimental Setup: Alice and Bob's measurement and detection setups are shown. Alice pumps a BBO crystal with a UV diode laser at 405~nm resulting in $810$~nm pairs of single photons through SPDC, and the pump is filtered out using a $10$~nm width bandpass filter centred at $810$~nm. The entangled pairs are separated using a knife edge mirror as shown in inlay~\textbf{a}. Entangled pairs are anti-correlated in their momentum and thus one photon from each pair falls on each side of the knife edge, as shown in the inlay. The single photons are coupled to a single mode fibre (SMF). The idler photon is then sent directly to Bob and the signal photon is sent to Alice's state preparation stage which uses a spatial light modulator (SLM) to prepare the OAM carrying beams. A diffraction grating is applied to the phase hologram which results in the formation of the desired mode in the first order of diffraction. A 4-f lens system with a pinhole at the focus then removes all the diffraction orders other than the first order. The holograms used for the dimension 4 states along with the corresponding modes is shown in inlay~\textbf{b}. Bob measures the incoming photon's state using an SLM and SMF, again using a 4-f system to remove all of the diffraction orders other than the first. The collected photons and idler photons are detected using single-photon avalanche diode (SPAD) detectors which then trigger a coincidence box to measure the coincidences.}
		\label{fig:setup}
	\end{center}
\end{figure*}
\noindent characteristic property of having an azimuthally dependent phase $\braket{\mathbf{r}}{\ell}:=e^{i \ell \phi}/\sqrt{2\pi}$, where $\ell$ is an integer from $-\infty$ to $+\infty$, and $\phi$ is the azimuthal angle in the polar coordinates $\mathbf{r}$. Such photonic states carry angular momentum of magnitude $\ell\hbar$ per photon along propagation direction. These photonic quantum states form complete and orthogonal bases, which we represent with $\ket{\ell}$. The OAM states for $\ell = \left\{ -4, -3, \ldots, 3, 4\right\}$, with the computational and the corresponding Hadamard states, are shown in Fig.~\ref{fig:modes}. In two dimensions the protocol is similar to that used in BB84 QKD. One takes first the computational basis given by $\left\{\ket{\psi_0}:=\ket{-1}; \, \ket{\psi_1}:=\ket{+1}\right\}$.  The conjugate basis is then taken as the Hadamard,
$\left\{\ket{\phi_1}:=(\ket{-1}+i\ket{+1})/\sqrt{2};\ket{\phi_2}:=(\ket{-1}-i\ket{+1})/\sqrt{2}\right\}$. As these states form a pair of MUBs, a projective measurement of a state in the correct basis gives a certain result, due to the orthogonality of the states in each basis, while measurement in the wrong basis gives no information, or a $50\%$ probability of detection for each state in the 2-dimensional case. When we move to higher dimensions we again will use two MUBs. However, each MUB will now have $d$ different states. In the higher dimensions, the states for the Hadamard basis are given by $\ket{\phi_i}=\sum_{j=1}^d e^{i j\, \pi /d}\ket{\psi_j}$. For dimension 4, the states would be $\ell = \{-2,-1,+1,+2\}$ and for dimension~8, the states would be $\ell = \{-4,-3,-2,-1,+1,+2,+3,+4\}$. More generally, the computational basis will be the pure OAM states from $\ell\in\left\{-d/2,  ... 0, ... ,d/2\right\}$ for the odd dimensions, and $\ell\in\left\{-(d-1)/2, \ldots,-1,1,\ldots ,(d-1)/2\right\}$ excluding $\ell=0$ for even $d$.

\noindent\textit{Experiment --} The experimental setup consists of a single photon source and OAM state preparation for Alice, and an OAM measurement system for Bob. Single photons are produced by pumping a BBO crystal with a 405~$nm$ UV laser. Degenerate photon pairs are selected using a 10~$nm$ bandpass filter centred at 810~$nm$ wavelength. A knife edge mirror is placed after the filter to separate the photon pairs which are anti-correlated in momentum. Each of these halves of the beam are then coupled to a single mode fibre where the signal photon is sent to Alice's state preparation system and the idler photon is sent to a detector. When coupled to the single mode fibres, the source has a rate of approximately 22~KHz, which reduces to 1 KHz after propagation through the experimental setup. The losses are mostly due to the spatial light modulators (SLMs) used for state preparation and detection. Alice uses a SLM to generate her desired OAM states. SLMs are not 100\% efficient, thus a holography approach is used, adding a diffraction grating to the phase mask which results in the desired OAM mode being formed in the 1st order of diffraction with very high state purity. After the SLM a 4-f lens system with a pinhole at the focus is used to filter out all diffraction orders except for the first order.

Bob also has an SLM and uses the phase flattening technique to measure the states sent by Alice. This technique involves applying the conjugate phase of the OAM mode that is to be measured and can be seen as removing any transverse phase from the incoming OAM carrying photon. This allows the photon to be coupled to a single mode fibre upon propagation, as the flat phase corresponds to a Gaussian mode in the far field. It must be stated that this is a projective measurement in which Bob must choose which state he will measure and only measures a single state at a time, thus limiting the efficiency of this particular measurement technique. There do however exist different approaches which can sort OAM modes upon choosing the basis in which one wants to measure. At present, many of these approaches suffer from efficiency problems which limit the eventual transmission rates. The idler photon is sent to Bob to perform a coincidence measurement with the signal photon, reducing noise from background light and detector dark counts. Both photons are detected using single-photon avalanche diode (SPAD) detectors. Bob makes all measurements, in the computational and Hadamard basis, using the SLM with different phase patterns. 

\begin{figure*}[!htb]
	\begin{center}
		\includegraphics[width=2.0\columnwidth]{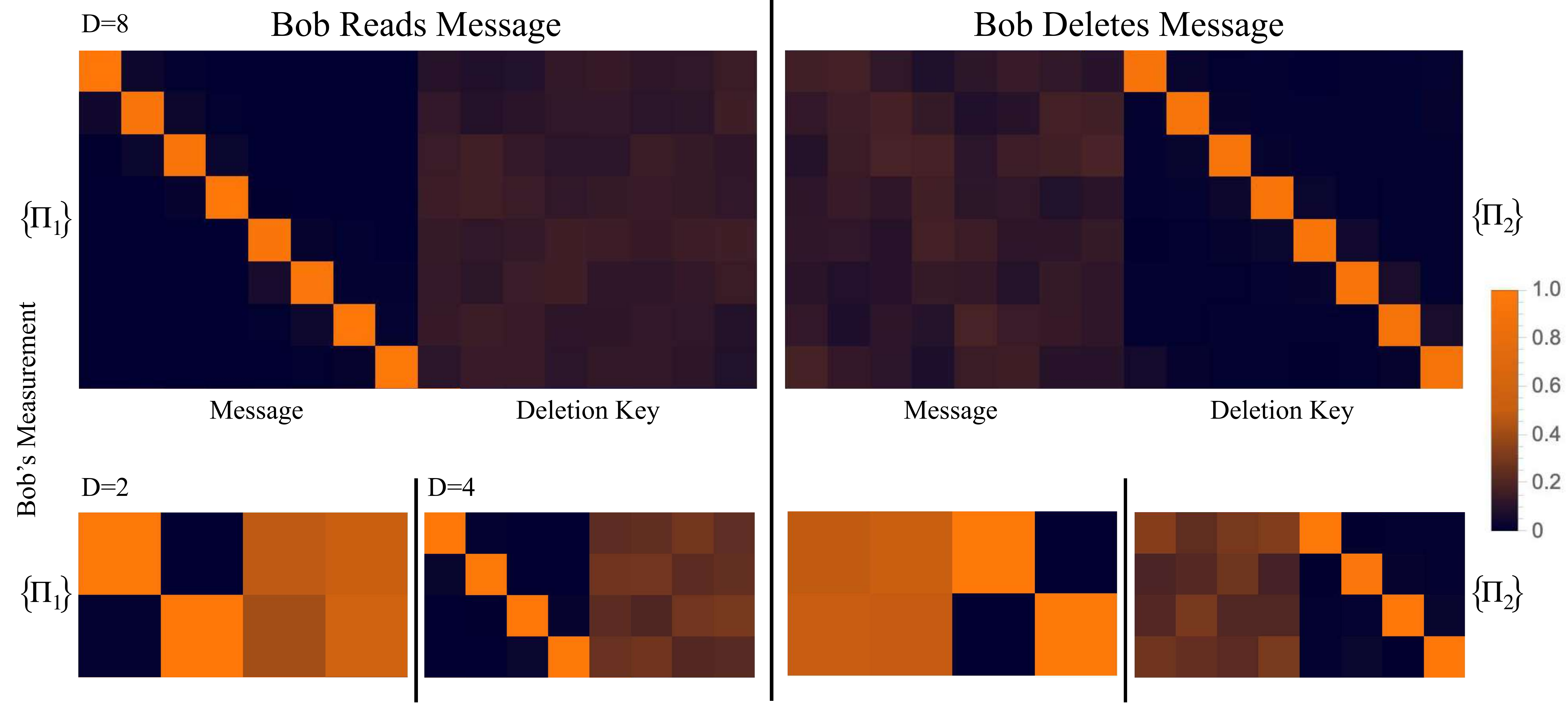}
		\caption{Probability of detection for dimension $d=2,4,$ and $8$: Bob's choice of measurement is shown in the rows of the detection matrix while the columns correspond to Alice's prepared state. On the left, Bob chooses to measure in the logical basis and thus reads the message sent by Alice. The states sent by Alice in the Hadamard basis are projected onto the logical basis and thus give no information, resulting in the uniform $1/d$ probability of detection. On the right, Bob measures in the Hadamard basis, thus erasing any information sent in the logical basis. The error rates observed are $\text{QBER}= 0.96\%, 2.4\%,$ and $7.2\%$ for dimension 2, 4, and 8 respectively, which correspond to a key rate per sifted photon of 0.84, 1.60, and 1.85.}
		\label{fig:d10}
	\end{center}
\end{figure*}

The experimental probability of detection matrices for the $d=2,4$ and $8$ QKD setup are shown in Fig.~\ref{fig:d10}. The columns correspond to the different states sent by Alice, while the rows correspond to the choice of measurement made by Bob. Thus we expect to find 100\% detection probability on the diagonal elements where Bob's measurement setting is the same state as that sent by Alice. When Bob measures Alice's states in the incorrect basis, he can gain no information and we expect to see a uniform probability of $1/d$ for all measurements in the wrong basis. When Bob reads the message, he performs all measurements in the logical basis, giving the measurement results shown on the left. This results in no information being gained about the deletion key, which can be seen by the incoherence of Alice's deletion key states. When Bob chooses to delete the message, he will instead choose to measure every state using the Hadamard basis, giving the results shown on the right. In this case we can see that the message states sent by Alice do not give any information. The error rates obtained are $\text{QBER}= 0.96\%, 2.4\%,$ and $7.2\%$ for dimension 2, 4, and 8, respectively. We can calculate the achievable message rate from our QBER ($Q$) using $K^{(d)}(Q)= \text{log}_2(d)-2\,h^{(d)}(Q)$ where the $d$-dimensional Shannon entropy is given by $h^{(d)}(x)= -x\, \text{log}_2(x/ (d-1))-(1-x)\,\text{log}_2(1-x)$. We have obtained message rates of 0.84, 1.60, and 1.85 per message photon for dimension 2, 4, and 8, respectively. Here, we see the advantage that can be achieved by moving to higher dimensional protocols, as the message rates here increase as the dimension of the protocol increases. 
\newline

\noindent\textit{Discussion --} A practical quantum information processing hardware called quantum memory, capable of arbitrary storing/releasing quantum states, has not yet been fully realised. Though the deletion concept may seem incomplete without a quantum memory, we can still conceive of other near-term applications in which Bob decides when the information is transmitted and whether he will keep or delete the information. For instance, Alice may continually send out an encryption key or software license, for which Bob must continue verifying that he has deleted until the time he would like to use it. Therefore, providing schemes to confirm a message has not been read will be vital for a quantum communication network. In this work, we have experimentally demonstrated the certified deletion protocol for a qubit QKD system as proposed by Broadbent and Islam~\cite{broadbent2020quantum}. We have also extended the protocol to include high-dimensional quantum states and have demonstrated this high-dimensional protocol, gaining an advantage in the message rate per sifted photon. The increase in message rate at higher dimensions also comes with a higher tolerance for errors (and therefore noise), which can be valuable for establishing communication in certain noisy environments. Another property of high dimensional states is that there are more than two MUBs that can be used to encode information. In fact, in dimensions where $d$ is a power of a prime number, \emph{i.e.}, $d=2,3,4,5,8, \ldots$, there exists $(d+1)$-MUBs. Unique QKD protocols, \emph{e.g.} six-state~\cite{bechmann1999incoherent}, using these extra MUB have been created that have additional key rate and security benefits over BB84. For certified deletion, one could come up with interesting new protocols involving multiple parties with messages or deletion keys encoded in the different bases. 
\\

\noindent This work was supported by Canada Research Chairs; Canada First Research Excellence Fund (CFREF); National Research Council of Canada High-Throughput and Secure Networks (HTSN) Challenge Program; and Natural Sciences and Engineering Research Council of Canada (NSERC).

\bibliographystyle{naturemag}
\bibliography{certifiedDeletion}

\providecommand{\noopsort}[1]{}
\begin{thebibliography}{10}
\expandafter\ifx\csname url\endcsname\relax
  \def\url#1{\texttt{#1}}\fi
\expandafter\ifx\csname urlprefix\endcsname\relax\def\urlprefix{URL }\fi
\providecommand{\bibinfo}[2]{#2}
\providecommand{\eprint}[2][]{\url{#2}}

\bibitem{park1970concept}
\bibinfo{author}{Park, J.~L.}
\newblock \bibinfo{title}{The concept of transition in quantum mechanics}.
\newblock \emph{\bibinfo{journal}{Foundations of physics}}
  \textbf{\bibinfo{volume}{1}}, \bibinfo{pages}{23--33} (\bibinfo{year}{1970}).

\bibitem{wootters1982single}
\bibinfo{author}{Wootters, W.~K.} \& \bibinfo{author}{Zurek, W.~H.}
\newblock \bibinfo{title}{A single quantum cannot be cloned}.
\newblock \emph{\bibinfo{journal}{Nature}} \textbf{\bibinfo{volume}{299}},
  \bibinfo{pages}{802--803} (\bibinfo{year}{1982}).

\bibitem{bennett1984quantum}
\bibinfo{author}{Bennett~Ch, H.} \& \bibinfo{author}{Brassard, G.}
\newblock \bibinfo{title}{Quantum cryptography: public key distribution and
  coin tossing int} \bibinfo{pages}{175--9} (\bibinfo{year}{1984}).

\bibitem{broadbent2009universal}
\bibinfo{author}{Broadbent, A.}, \bibinfo{author}{Fitzsimons, J.} \&
  \bibinfo{author}{Kashefi, E.}
\newblock \bibinfo{title}{Universal blind quantum computation}.
\newblock In \emph{\bibinfo{booktitle}{2009 50th Annual IEEE Symposium on
  Foundations of Computer Science}}, \bibinfo{pages}{517--526}
  (\bibinfo{organization}{IEEE}, \bibinfo{year}{2009}).

\bibitem{sit:18}
\bibinfo{author}{Sit, A.} \emph{et~al.}
\newblock \bibinfo{title}{Quantum cryptography with structured photons through
  a vortex fiber}.
\newblock \emph{\bibinfo{journal}{Optics Letters}}
  \textbf{\bibinfo{volume}{43}}, \bibinfo{pages}{4108--4111}
  (\bibinfo{year}{2018}).

\bibitem{rosenberg2007long}
\bibinfo{author}{Rosenberg, D.} \emph{et~al.}
\newblock \bibinfo{title}{Long-distance decoy-state quantum key distribution in
  optical fiber}.
\newblock \emph{\bibinfo{journal}{Physical review letters}}
  \textbf{\bibinfo{volume}{98}}, \bibinfo{pages}{010503}
  (\bibinfo{year}{2007}).

\bibitem{gobby2004quantum}
\bibinfo{author}{Gobby, C.}, \bibinfo{author}{Yuan, a.} \&
  \bibinfo{author}{Shields, A.}
\newblock \bibinfo{title}{Quantum key distribution over 122 km of standard
  telecom fiber}.
\newblock \emph{\bibinfo{journal}{Applied Physics Letters}}
  \textbf{\bibinfo{volume}{84}}, \bibinfo{pages}{3762--3764}
  (\bibinfo{year}{2004}).

\bibitem{bouchard2018underwater}
\bibinfo{author}{Bouchard, F.} \emph{et~al.}
\newblock \bibinfo{title}{Quantum cryptography with twisted photons through an
  outdoor underwater channel}.
\newblock \emph{\bibinfo{journal}{Optics express}}
  \textbf{\bibinfo{volume}{26}}, \bibinfo{pages}{22563--22573}
  (\bibinfo{year}{2018}).

\bibitem{Sit:17}
\bibinfo{author}{Sit, A.} \emph{et~al.}
\newblock \bibinfo{title}{High-dimensional intracity quantum cryptography with
  structured photons}.
\newblock \emph{\bibinfo{journal}{Optica}} \textbf{\bibinfo{volume}{4}},
  \bibinfo{pages}{1006--1010} (\bibinfo{year}{2017}).
\newblock
  \urlprefix\url{http://www.osapublishing.org/optica/abstract.cfm?URI=optica-4-9-1006}.

\bibitem{schmitt2007experimental}
\bibinfo{author}{Schmitt-Manderbach, T.} \emph{et~al.}
\newblock \bibinfo{title}{Experimental demonstration of free-space decoy-state
  quantum key distribution over 144 km}.
\newblock \emph{\bibinfo{journal}{Physical Review Letters}}
  \textbf{\bibinfo{volume}{98}}, \bibinfo{pages}{010504}
  (\bibinfo{year}{2007}).

\bibitem{Yin:2017}
\bibinfo{author}{Yin, J.} \emph{et~al.}
\newblock \bibinfo{title}{{Satellite-based entanglement distribution over 1200
  kilometers}}.
\newblock \emph{\bibinfo{journal}{Science}} \textbf{\bibinfo{volume}{356}},
  \bibinfo{pages}{1140--1144} (\bibinfo{year}{2017}).
\newblock
  \urlprefix\url{http://www.sciencemag.org/lookup/doi/10.1126/science.aan3211}.

\bibitem{broadbent2020quantum}
\bibinfo{author}{Broadbent, A.} \& \bibinfo{author}{Islam, R.}
\newblock \bibinfo{title}{Quantum encryption with certified deletion}.
\newblock In \emph{\bibinfo{booktitle}{Theory of Cryptography Conference}},
  \bibinfo{pages}{92--122} (\bibinfo{organization}{Springer},
  \bibinfo{year}{2020}).

\bibitem{garg2020formalizing}
\bibinfo{author}{Garg, S.}, \bibinfo{author}{Goldwasser, S.} \&
  \bibinfo{author}{Vasudevan, P.~N.}
\newblock \bibinfo{title}{Formalizing data deletion in the context of the right
  to be forgotten}.
\newblock In \emph{\bibinfo{booktitle}{Annual International Conference on the
  Theory and Applications of Cryptographic Techniques}},
  \bibinfo{pages}{373--402} (\bibinfo{organization}{Springer},
  \bibinfo{year}{2020}).

\bibitem{poremba2022quantum}
\bibinfo{author}{Poremba, A.}
\newblock \bibinfo{title}{Quantum proofs of deletion for learning with errors}.
\newblock \emph{\bibinfo{journal}{arXiv preprint arXiv:2203.01610}}
  (\bibinfo{year}{2022}).

\bibitem{bartusek2022cryptography}
\bibinfo{author}{Bartusek, J.} \& \bibinfo{author}{Khurana, D.}
\newblock \bibinfo{title}{Cryptography with certified deletion}.
\newblock \emph{\bibinfo{journal}{arXiv preprint arXiv:2207.01754}}
  (\bibinfo{year}{2022}).

\bibitem{coiteux2019proving}
\bibinfo{author}{Coiteux-Roy, X.} \& \bibinfo{author}{Wolf, S.}
\newblock \bibinfo{title}{Proving erasure}.
\newblock In \emph{\bibinfo{booktitle}{2019 IEEE International Symposium on
  Information Theory (ISIT)}}, \bibinfo{pages}{832--836}
  (\bibinfo{organization}{IEEE}, \bibinfo{year}{2019}).

\bibitem{bechmann2000quantum}
\bibinfo{author}{Bechmann-Pasquinucci, H.} \& \bibinfo{author}{Tittel, W.}
\newblock \bibinfo{title}{Quantum cryptography using larger alphabets}.
\newblock \emph{\bibinfo{journal}{Physical Review A}}
  \textbf{\bibinfo{volume}{61}}, \bibinfo{pages}{062308}
  (\bibinfo{year}{2000}).

\bibitem{Cerf2002}
\bibinfo{author}{Cerf, N.~J.}, \bibinfo{author}{Bourennane, M.},
  \bibinfo{author}{Karlsson, A.} \& \bibinfo{author}{Gisin, N.}
\newblock \bibinfo{title}{{Security of Quantum Key Distribution Using d -Level
  Systems}}.
\newblock \emph{\bibinfo{journal}{Physical Review Letters}}
  \textbf{\bibinfo{volume}{88}}, \bibinfo{pages}{127902}
  (\bibinfo{year}{2002}).
\newblock
  \urlprefix\url{http://link.aps.org/doi/10.1103/PhysRevLett.88.127902}.
\newblock \eprint{0107130}.

\bibitem{csiszar1978broadcast}
\bibinfo{author}{Csisz{\'a}r, I.} \& \bibinfo{author}{Korner, J.}
\newblock \bibinfo{title}{Broadcast channels with confidential messages}.
\newblock \emph{\bibinfo{journal}{IEEE transactions on information theory}}
  \textbf{\bibinfo{volume}{24}}, \bibinfo{pages}{339--348}
  (\bibinfo{year}{1978}).

\bibitem{bechmann1999incoherent}
\bibinfo{author}{Bechmann-Pasquinucci, H.} \& \bibinfo{author}{Gisin, N.}
\newblock \bibinfo{title}{Incoherent and coherent eavesdropping in the
  six-state protocol of quantum cryptography}.
\newblock \emph{\bibinfo{journal}{Physical Review A}}
  \textbf{\bibinfo{volume}{59}}, \bibinfo{pages}{4238} (\bibinfo{year}{1999}).

\end{thebibliography}

\end{document}